\documentclass[preprint,aps,epsf]{revtex4}
\usepackage{graphicx}
\usepackage{epsf}
\usepackage{wrapfig}
\usepackage{epsfig}
\newcommand{\beq}{\begin{equation}}
\newcommand{\eeq}{\end{equation}}
\newcommand\ba{\begin{eqnarray}}
\newcommand\be{\begin{equation}}
\newcommand\ee{\end{equation}}
\newcommand\la{\left\langle}
\newcommand\ra{\right\rangle}
\newcommand\ea{\end{eqnarray}}
\begin{document}
\title{On the dependence of the wave function of a bound nucleon  on its
momentum  and the EMC effect}
\author{C. Ciofi degli Atti}
\address{Department of Physics\\ University of Perugia and
      INFN Sezione di Perugia,\\
      Via A. Pascoli, I-06123 Perugia, Italy}
\author{L.~L.~Frankfurt}
\address{School of Physics and Astronomy,
Tel Aviv University,
         69978 Tel Aviv, Israel}
\author{L. P. Kaptari}
\address{ Bogoliubov Lab.
      of Theoretical  Physics, 141980, JINR,  Dubna, Russia\\
      and\\ Department of Physics, University of Perugia and INFN
  Sezione di Perugia,
      Via A. Pascoli, I-06123 Perugia, Italy}
\author{M.~I.~Strikman}
\address{Department of Physics,
 Pennsylvania
State University, University Park, PA 16802, USA,\\
and\\ Deutsches Elektronen Synchrotron DESY, Germany
}
\date{\today}
\begin{abstract}
It is  widely discussed in the literature
  that the wave function of the nucleon bound in a
nucleus is modified due to the interaction with the surrounding
medium. We argue that  the modification should strongly depend on
the momentum of the nucleon. We study such an effect in the case
of the point-like configuration component of the wave function of
a nucleon bound in a nucleus $A$,  considering  the case of
arbitrary final state of the spectator $A-1$ system. We show that
for non relativistic values of the nucleon momentum, the  momentum
dependence of the nucleon deformation appears to follow from
rather   general  considerations and discuss the implications of
our theoretical observation for two different phenomena:
 i) the search for medium induced modifications of the nucleon radius
 of a
 bound nucleon through the measurement of the  electromagnetic nucleon
form factors via  the $A(e,e'p)X$ process, and ii) the
A-dependence of the EMC effect; in this latter case we also
present a new method of estimating the fraction of the nucleus
light-cone momentum carried by the
 photons  and find that in a heavy nuclei protons loose about 2\% of their momentum.

 \end{abstract}
\maketitle
\section{Introduction}

 One of the main  challenges in
particle physics, nuclear physics and astrophysics is the
necessity to achieve unambiguous understanding of the physics of
cold dense nuclear matter, the limiting mass and radius of neutron
stars, and many other questions which  deeply interconnect
astrophysics with the physics of particles and nuclei. To this
end,  it is important to reliably evaluate  the modifications of
the wave function of a nucleon embedded in  dense nuclear matter.
The laboratory investigation
 of the quark-gluon structure of a nucleon bound within a nucleus may
help to  quantify this important phenomenon.
 Twenty five
 years after the discovery of the suppression of the nucleus structure function
as compared to that for a free nucleon at moderate value of $x$-
the EMC effect (see e.g. Ref. \cite{EMC})-,  its origin is still a
matter of discussions and various effects have been advocated to
explain it, like, e.g.: (a) the modification of the  bound nucleon
structure function due to (i) possible change of non-perturbative
QCD scale in nuclei \cite{close}, (ii) the  meson - nucleon
interactions \cite{thomas}, (iii) the dependence of the strength
of the nucleon interaction upon  the size of the quark-gluon
configuration which leads to oscillations of this effect as the
function of nuclear density \cite{FS85}; (b) the  presence of
non-nucleonic dynamic degrees of freedom in the nucleus (see for
example \cite{pions}) which carry a fraction of the total nucleus
momentum,  leading to depletion from one of the fraction of
nucleus momentum carried by nucleons;  c) relativistic effects due
to nucleon binding and Fermi motion \cite{binding,Simo}, models
using the Bethe-Salpeter vertex function as nucleus wave function
\cite{relativistic}. Experiments at Jlab, especially after the 12
GeV upgrade, will be able to break the deadlock by performing a
series of dedicated experiments (see e.g. \cite{Jlab}).

Certain restriction on the models follows from the investigation
of Drell-Yan process \cite{DY} which found no enhancement of
antiquark distribution in nuclei.  It  appears difficult to
explain this fact within models  where non-nucleonic degrees of
freedom are mesons. The conclusion, which follows solely from the
requirements of the baryon charge and momentum conservation, is
that the EMC effect signals the presence of non-nucleonic degrees
of freedom in nuclei, though it is not yet clear which are the
most relevant ones.

A possibility,  which is  discussed in a number of models of the
nucleon,  is that the bound nucleon wave function is deformed due
to the presence of nearby nucleons. A distinctive property of QCD,
which is a consequence of color gauge invariance, is that
different components of the hadron wave function interact with
different strengths.
 The extreme example is the point-like
configurations (PLC) in hadrons which have interaction strength
much smaller than the average one, leading to the phenomenon of
color transparency (for a recent review see e.g.
\cite{Stanreview}).
%in small $x$ processes \cite{Gunion-Soper,FS85}
%and in moderate $x$ processes \cite{color}.
%Possible effects of color
%transparency in different processes have been suggested in various
%papers \cite{Brodsky, FS88,Boris, Bertsch, Nikolaev}.
%\cite{Brodsky:1988xz,FS88,Boris,...}
This phenomenon has been recently observed for the process of
coherent high-energy pion dissociation into two jets
\cite{Ashery}, with various characteristics of the process
consistent with the original QCD predictions\cite{FMS93}.  There
are also evidences for CT effects in  quasielastic production of
$\rho$ mesons off nuclei, investigated firstly by the  E665
experiment at FNAL \cite{Adams}
%\cite{Adams:1994bw}
 and, recently, by
the  HERMES experiment at DESY \cite{HERMESRHO}. The data of
HERMES are well described  by the model \cite{Borisrho} which
takes into account  the squeezing of the $q\bar q$ state in the
production vertex and its expansion while propagating through the
nucleus.

We will consider in this paper the effect of the suppression of
PLC in bound
 nucleons,  extending
the analysis  to the case of a nucleon bound in $^3He$ and in
complex nuclei,  with the spectator  system being in  specific
energy states. Similar to \cite{FS85,FS88} we find that the effect
of suppression of PLC's strongly depends on the momentum of the
nucleon. Moreover, by taking into account the energy state of the
spectator $A-1$ system,   we find a new effect, namely a strong
dependence of the effect on the excitation energy of the residual
system. Overall the analysis of the derived formula allows us to
establish the connection with another language, which explores the
concept of  off mass shell particles. As a matter of fact,  we
find that the effect depends on the virtuality of the interacting
nucleon defined via the kinematics  of the spectator system.

More generally, we  will argue that a  strong dependence of the
deformation of the nucleon wave function upon  the momentum, is a
general phenomenon in the lowest order over $p^2/\mu^2$, where $p$
is the nucleon momentum and $\mu \sim 0.5 \div 1\,  GeV $ is the
strong interaction scale. We also discuss two implications of this
argument; the first one is the need to look for the deviations of
the bound nucleon electromagnetic (e.m.) form factors from the
free ones as a function of the struck nucleon momentum; the second
one is the estimate the A-dependence of the EMC effect through the
energy binding and the mean excitation energies of nuclei. Our
paper is organized as follows: in Section II we briefly review the
arguments concerning the reduction of PLC in the nuclear medium
and extend the analysis of Ref. \cite{FS85,FS88} to three- and
many-body  nuclei; the connection between nucleon virtuality and
the suppression of PLC's is presented in Section III; the effects
of the reduction of PLC's in Quasi Elastic and Deep Inelastic
scattering
 is discussed in Section IV; in Section V the results of our
 calculations of the EMC effect in nuclei, including the deuteron,
 are presented;
 in Section VI the contribution of the coherent Weizecker-Williams
 photon field of the nucleus and its implications for the EMC effects in heavy nuclei are discussed;
 eventually, the Conclusions are given in Section VII.

\section{The suppression of Point-Like Configurations in nuclei}
\label{sec1}
We will consider how the wave function of a bound nucleon is modified due to
 medium effects. Our approach
 implements a well understood and established property of pQCD: if the collision energy
 is not too large,  the interaction between hadrons is proportional to their size.
 This property
 relies on the non relativistic Schr\"odinger equation
 for the  nucleus wave function  which describes the motion of centers of mass  of the nucleons.
  Some notations are therefore  in order.
 The  Schr\"odinger equation for a nucleus composed of $A$ nucleons
 interacting via two-body interactions, is
 \beq
 H_A\Psi_A^f = \left[\sum_i \frac{\bf p_i^2}{2m_N} +\sum_{i<j}V_{ij}
 \right] \Psi_A^f= E_A^f \Psi_A^f,
 \label{schroedinger}
 \eeq
 where
 the index
 $f\equiv \{0,1,2,...\}$ denotes the excitation spectrum of the system
  (note that from now-on  the ground-state energy and wave functions
 will be simply denoted by $E_A$ and $\Psi_A$, instead of $E_A^{(0)}$ and
 $\Psi_A^{(0)}$;
 moreover, in case of the Deuteron, instead of $A=2$ we will simply use the notation $D$).
  We will consider a nucleon with four momentum $p \equiv(E_p,{\bf p})$
 and denote the Center-of-Mass  four momentum of the  {\it spectator}
 $(A-1)$ nucleons as $p_s$ or $P_{A-1}$.
  The mass of the nucleon will be denoted by $m_N$. We will also need
  to define the energy necessary
 to remove a nucleon from a nucleus $A$,  leaving the residual system in
  a state with intrinsic (positive) excitation energy $E_{A-1}^f$;
  such a quantity  is the (positive )
  {\it removal energy} defined as $E=E_{min} + E_{A-1}^f$ with
 $E_{min}= |E_A| -|E_{A-1}|$,   $E_A$ and $E_{A-1}$ being  the
 (negative) ground-state energies of $A$ and $A-1$ systems. Eventually,
the  ground-state  energy per particle will be denoted by
$\epsilon_A =E_A/A$.

\subsection{General considerations}

In QCD the Fock space decomposition of the hadron wave function
contains components of the size much smaller than the average size
of the hadron;  these components determine, at $Q^2\to \infty$,
the asymptotic behaviour of the elastic hadron form factors and,
for a pion,  they were explicitly  observed in the exclusive dijet
production \cite {Ashery}.

It was argued in Ref.\cite{FS85,FS88} that since the small size
configurations of the bound nucleon experience a smaller nucleon
attraction, their probability should be smaller in the bound state
since such a reduction would lead  back to an increase of the
nuclear binding. The discussed effect was  formally described by
an expression obtained within the closure approximation
\cite{FS85,FS88,Miller}.

The reduction of PLCs  might be  relevant for the explanation of
the EMC effect, though only in a restricted  region of the Bjorken
scaling variable $x=\frac{Q^2}{2m_N\nu}$. Indeed, it  has been
predicted by several models, see e.g.  \cite{small,FS85} that the
behavior of the structure functions at $x\to 1$ should be
sensitive to the small size  quark gluon configurations. One
general argument is that at moderate values of the four-momentum
transfer $Q^2$, PLCs compete in elastic form factors with end
point contribution which is also but gradually squized with
increase of $Q^2$ as the consequence of Sudakov type form factors,
which, on the other side, are connected to the  inclusive
structure functions at $x\to 1 $  via   the Drell-Yan-West
relation (see the discussion in \cite{Drell-Yan,Gribov-Lipatov}).
Another argument, mostly relevant for the nucleon parton density,
is that large size configurations with the pion cloud do not
contribute at  $x \simeq 1$,  since in these configurations pions
carry a significant part of the total light-cone momentum of the
nucleon.

The key characteristic of PLCs,   which allowed one to derive a
compact expression for their  modification in the bound nucleon,
is that
 the potential energy associated to the interaction of a  PLC
 is much smaller than  $V_{ij}$, the
 NN potential averaged over {\it all } configurations.
 Using the decomposition of  the PLC over the hadronic states and the  closure approximation
 one finds  for the nuclear wave function which includes PLC
 in the nucleon $i$, the following expression~\cite{FS85,FS88}
 \begin{eqnarray}
\tilde \psi_A(i)
 \approx \left (1+ \sum_{j\ne i}\frac{V_{ij}}{\Delta E^{(N/A)}} \right )
 \psi_A(i),
\label{dpsi}
\end{eqnarray}
where $\psi_A(i)$ is the usual wave function with all nucleons,
including $i$,
in average configurations and
$\Delta E^{(N/A)} \sim m_{N^*}-m_N\sim 600-800  MeV$  parametrizes
the energy denominator depending upon  the average virtual excitation
of a nucleon $N$ in the nucleus $A$.

 \subsection{The deuteron}

Using the equations of motion for $\psi_A$,  the momentum
dependence of the probability to find a bound nucleon with momentum
$p$ in a PLC  was obtained   in Ref. \cite{FS85,FS88}  within the
 mean field and two nucleon correlation approximations. In particular,
 for the deuteron  the Shr\"odinger equation   in momentum  representation leads to
\begin{eqnarray}
\sum_{j\ne i}{V_{ij}}  = -2\frac{\bf {p}^2}{2m_N} + E_D,
\label{dpsid}
\end{eqnarray}
where $E_D$ is the (negative) binding energy of the deuteron.
Using the same closure approximation  and equation of motions for the higher order terms
in $\frac{V_{ij}}{\Delta E}$ (assuming that $\Delta E$ is approximately the same
for the higher order terms) one obtains  that  the suppression of
the probability of PLC's in the deuteron for a nucleon with momentum ${\bf p}$
is given by
\begin{eqnarray}
\delta_{D}({\bf {p}}) =
\left (1+\frac{2\frac{\mbox{${\bf{p}}^2$}}{\mbox{$2m_N$}} - E_D}
{\Delta E^{(N/D)}}
\right )^{-2}.
\label{deldeapp}
\end{eqnarray}

Thus, if for a given $x$   PLCs   dominate in the nucleon Parton
Distribution Functions, the structure function of the bound
nucleon would be suppressed by a factor given by
 Eq. (\ref{deldeapp}), that is
\begin{eqnarray}
F_D (x,{\bf {p}},Q^2) \simeq  \delta_{D}({\bf {p}}) F_{2N}(x,Q^2)
\simeq \left (1+\frac{2\frac{\mbox{${\bf{p}}^2$}}{\mbox{$2m_N$}} -
E_D}{\Delta E^{(N/D)}} \right )^{-2} F_{2N}(x, Q^2).
 \label{delde}
\end{eqnarray}
 Note that  Eq.(\ref{delde}) can equally well be applied to the semi-inclusive process
 when the transition to one particular
 final state of the spectator is considered.
In the derivation of the previous formulae, it has been assumed that for PLC
 $|V_{ij}^{(SC)}(x)|/|V_{ij}| \ll 1$. If one probes large values of $x$,
 for which $|V_{ij}^{(SC)}(x)|/|V_{ij}| =\lambda(x) <1$, the suppression factor will
  be obviously smaller and,   in the lowest order in $p^2/2m_N \Delta E$,
   Eq. (\ref{deldeapp}) will be modified as follows

 \begin{eqnarray}
\delta_{D}({\bf {p}}) =
\left (1+[1-\lambda(x)]\frac{2\frac{\mbox{${\bf{p}}^2$}}{\mbox{$2m_N$}} - E_D}
{\Delta E^{(N/D)}}
\right )^{-2}.
\label{deldeapp1}
\end{eqnarray}

 If   the dominant nucleon configurations in $F_{2N}$ interact with a strength
 substantially smaller than the average (say $\lambda\le 0.5$)
    for $x\ge 0.5\div 0.6$,  the PLC  suppression may help explaining   the EMC effect.
Since we are interested in this paper in the $A$-dependence of the
deviation of the  EMC ratio from one, for ease of presentation, we
will simply put in what follow $\lambda (x) = 0$, though in the
comparison with experimental data, to be presented in Section V, a
value of $\lambda(x)\neq 0$ has been used.

 A key test of the PLC suppression  is the study of the tagged structure functions
 \cite{FS85,MSS,simula,CKS,CKK,sargisianFS}.

\subsection{The three-body nuclei}
Let us consider now the three-body nuclei. In this case we face  a more complicated situation
due  to  several possible final states of the two-body spectator system. For this reason,
the suppression  of PLCs will depend upon the transition
densities between the wave function (\ref{dpsi}) and the final state wave functions. Therefore
 the full nuclear Spectral  Function of
$^3He$  is required to evaluate $\delta_{^3He}$.  Let us discuss this point in details. To this end,
 we first introduce the   wave function $\phi_2$,  the
solution of the two-body Schr\"odinger
equation for nucleons $"2"$ and $"3"$:
\begin{eqnarray}
(\hat T_2 +\hat T_3 + V_{2,3})\phi_2^f(2,3) = E_2^f \phi_2^f(2,3),
\label{sh2}
\end{eqnarray}
where $T$ is the operator for the kinetic energy,
 $f$ labels the quantum numbers of the state which can be  either the ground $(D)$,
or the continuum $(pn)$ states of a neutron-proton pair,
 or the continuum $(nn)$ state of a neutron-neutron pair
(note, that Eq. (\ref{sh2}) has the same spectrum as
the final two-nucleon state in the case of deep inelastic scattering on $^3He$, i.e. $D$, and $pn$
and $pp$ in the continuum). Then
the relevant quantities are the following densities
\begin{eqnarray}
\phi_1^\dagger(1)\phi_2^{f\dagger}(2,3)\tilde \psi_{3}(1,2,3)
 = \phi_1^\dagger(1)\phi_2^{f\dagger}(2,3)
 \left (1+  \frac{V_{1,2}+V_{1,3}}{\Delta E^{(N/3)}}\right )\psi_{3}(1,2,3),
\label{denhe}
\end{eqnarray}
where $\phi_1(1)$ is the wave function of the struck nucleon.
By considering the full three-nucleon Shr\"odinger equation
\begin{eqnarray}
(\hat T_1 +\hat T_2 +\hat T_3 + V_{2,3} + V_{1,3}+ V_{1,2})\psi_{3}(1,2,3) =
 E_{3} \psi_{3}(1,2,3),
\label{sh3}
\end{eqnarray}
we obtain
\begin{eqnarray}
\phi_1^{\dagger}(1)\phi_2^{f\dagger}(2,3)( V_{1,3}+ V_{1,2})\psi_{3}(1,2,3) =
( E_{3} - E_2^f - T_1)\phi_1^{\dagger}(1)\phi_2^{f\dagger}(2,3)\psi_{3}(1,2,3).
\label{sh3p}
\end{eqnarray}
The density in r.h.s. of this equation  defines the channels $f$
of the spectral function of $^3He$, to be denoted    $P_3^{(f)}(|{\bf{p}}|,E)$ ~\cite{CPS}.
For the three different channels we obtain
(in case of the inclusive process, there is a sum over states of the
spectator, $f$):
\begin{eqnarray}
&& \delta_{3}^ {(D)}({\bf {p}})
 \simeq
 \left (1+\frac{|E_3|- |E_D| +\frac{\mbox{$3{\bf p}^2$}}{\mbox{$4m_N$}}}{\Delta E^{(N/3)}}
 \right )^{-2} ,
\label{delpd}\\
&& \delta_3^{(pn)}({\bf{p}},{\bf{k}})
 \simeq
 \left  (1+
\frac
{
|E_3|+ \frac{\mbox{${\bf k}^2$}}{\mbox{$m_N$}}+
\frac{\mbox{$3{\bf{p}}^2$}}{\mbox{$4m_N$}}
}
 {\Delta E^{(N/3)}}
 \right )^{-2} ,
\label{delppn}\\
&& \delta_{3}^{(pp)}({\bf{p}},{\bf{k}})
 \simeq
 \left  (1+
\frac
{
|E_3|+ \frac{\mbox{${\bf k}^2$}}{\mbox{$m_N$}}+
\frac{\mbox{$3{\bf{p}}^2$}}{\mbox{$4m_N$}}
}
 {\Delta E^{(N/3)}}
 \right )^{-2},
\label{delnpp}
\end{eqnarray}
%\label{delnpp}
%\end{eqnarray}
where we neglected the difference of the proton and neutron masses and also
 a possible isospin dependence of $\Delta E^{(N/3)}$.
Here ${\bf k}$ is the  momentum of the nucleon in the spectator pair in the pair's c.m.
 frame. In terms of {\it removal energies} we have $E_{min}=|E_3| - |E_2|$
  for the process
 $^3He \rightarrow p+D$, and  $E_{min}=|E_3|$ for the
  process  $^3He \rightarrow p + (pn)$; the corresponding excitation
  energies are
   $E_{A-1}^f=0$ and  $E_{A-1}^f={\bf k}^2/m_N$, respectively.
   Thus Eqs. (\ref{delpd}), (\ref{delppn}), and
   (\ref{delnpp}) can be unified as
   \begin{eqnarray}
\delta_{3}^{(f)}({\bf{p}},E)
 \simeq
 \left  (1+
\frac
{E +\frac{\mbox{$3{\bf p}^2$}}{\mbox{$4m_N$}}}
{\Delta E^{(N/3)}}
 \right )^{-2},
\label{delge3}
\end{eqnarray}
where $E=E_{min}+E_2^f$ generates the dependence upon $f$ of the
r.h.s.. We will need in what follows the average value of
$\delta_{3}^{(f)}({\bf{p}},E)$ with respect to ${\bf{p}}$ and $E$.
This can be obtained provided the nucleon spectral function in
channel $f$, $P_3^{(f)}(|{\bf{p}}|,E)$, is known; in such a case
one has \beq \left< \delta_{3}^{(f)}({\bf{p}},E)\right> =
 \int \delta_{3}^{(f)}({\bf{p}},E)P_3^{(f)}(|{\bf p}|,E)\,dE d{\bf
 p} .
\label{average} \eeq

Following Ref. \cite{CPS}, we will label various quantities
pertaining to the channel  $^3He\rightarrow D +p$ with the superscript
 (${\it gr}$),
and
quantities pertaining to the channels  $^3He\rightarrow p+(np)$
and  $^3He\rightarrow n +(pp)$ with the superscript
 ${\it ex}$.
The corresponding Spectral Functions will be denoted $P_3^{(D)}({\bf{p}},E) \equiv P_3^{(gr)}({\bf{p}},E)$
and $ P_3^{(NN)}({\bf{p}},E) \equiv P_3^{(ex)}({\bf{p}},E)$.

 The Spectral Functions in different channels $f$ are normalized as following
 \be
 \int P_3^{(f)}(|{\bf p}|,E) d\,{\bf p}\,\,dE = S_f ,
 \label{norm}
 \ee\\
 where $f = \{gr,ex\}$.

 The spectral function of $^3He$ is described in details in Appendix A
  and the  mean values of various quantities calculated with a realistic Spectral Function
  of  $^3He$
 \cite{leonya3} are listed in Table 1.

\subsection{General case}
Eq. (\ref{delge3}) can be readily generalized to  the case of an arbitrary  nucleus $A$, obtaining
\begin{eqnarray}
\delta_{A}^{(f)} ({\bf{p}},E)
 \simeq
 \left  (1+
\frac
{E + \frac  {\mbox {$A$}}{\mbox{$A-1$}}   \frac{\mbox{${\bf p}^2$}}
      {\mbox{$2M$}}}
{\Delta E^{(N/A)}}
 \right )^{-2}.
\label{delgenerale}
\end{eqnarray}
It is also trivial to modify Eq.(\ref{delgenerale})  to account
for the case of small but finite size configurations by
introducing a factor $1-\lambda(x)$ as in Eq.(\ref{deldeapp1}).
The evaluation of the average values of Eq. (\ref{delgenerale})

\beq \left< \delta_{A}^{(f)}({\bf{p}},E)\right>= \int
\delta_{A}^{(f)}({\bf{p}},E)P_A^{(f)}(|{\bf p}|,E)\,dE d{\bf p},
\label{average1} \eeq \noindent  requires  the knowledge of the
Spectral Function of the nucleus $A$ in channel $f$,
$P_A^{(f)}(|{\bf p}|, E)$. As illustrated in Refs. \cite{Simo,CS},
the Spectral Function of a Complex Nucleus can be written in the
following form \beq P_A(|{\bf p}|,E)=P_{0}(|{\bf
p}|,E)+P_{1}(|{\bf p}|,E) , \label{spectralA} \eeq where
$P_{0}(|{\bf p}|,E)$ describes the transition to the ground state
and to the discrete shell-model states of the nucleus $A-1$,
whereas
  $P_{1}(|{\bf p}|,E)$ is responsible for the transitions to
the whole of the continuum states generated by
short-range nucleon-nucleon
 correlations.
 %Numerical results will be presented in the subsequent sections.Therefore
 For complex nuclei we will consider two  average values
for the suppression of PLCs, namely

\begin{eqnarray}
\left < \delta_{A}^{(0)}({\bf p}, E) \right > \simeq \int
\delta_{A}^{(0)}({\bf{p}},E)P_{0}(|{\bf p}|,E)\,dE d{\bf p}
,\label{del0}
\end{eqnarray}

\begin{eqnarray}
\left < \delta_{A}^{(1)}({\bf p}, E) \right > \simeq \int
\delta_{A}^{(1)}({\bf{p}},E)P_{1}(|{\bf p}|,E)\,dE d{\bf p} ,
 \label{del1}
\end{eqnarray}
whre  $f=\{0,1\}$  plays, in a sense, the role of $f={gr,ex}$ in
the case of $^3He$. The mean values of various quantities
pertaining to complex nuclei calculated with the spectral function
of Ref. \cite{CS} are reported in Table 2.

\section{Nucleon virtuality, the suppression of PLCs  and the variation
of nucleon properties in the medium}

\subsection{The nucleon virtuality}
Let us  now consider the interaction of a bound nucleon with a virtual photon.
The virtuality of the interacting nucleon $v$ is as follows
\begin{equation}
v = p^2-m_N^2 = (P_A-P_{A-1})^2-m_N^2.
\label{virt}
\end{equation}

In impulse approximation (${\bf p}=-{\bf P}_{A-1}$) we have
\begin{eqnarray} &&
v(|{\bf p}|, E)=(P_A-P_{A-1})^2-m_N^2=(M_A - P_{A-1}^{(0)})\,\,^2
- {\bf p}^2 - m_N^2 =\nonumber \\&& =\left(M_A -\sqrt{(M_{A}-
m_N+E)^{2}+{\bf p}^{2}}\,\,\right)^2-{\bf p}^2 - m_N^2
.\label{rem}
\end{eqnarray}
The non-relativistic reduction of Eq.\,(\ref{rem})
in the rest frame of the nucleus $A$,
which corresponds to neglecting higher order terms in
$\sim\displaystyle\frac{E}{m_N}$ and  $\sim
\displaystyle\frac{T_{A-1}}{m_N}$, yields
\begin{eqnarray} &&
v_{NR}(|{\bf p}|, E) \approx -2m_N \left(\frac{A}{A-1}\frac{{\bf
p}^2}{2m_N} + E\right ).
 \label{rem1}
\end{eqnarray}

It can therefore be seen  that in the non-relativistic limit
the argument of $\delta_A(|{\bf p}|,E)$ for any $A$,
is the same as the non relativistic reduction  of the virtuality $v_{NR}$,
so that  the suppression of PLCs  can be expressed in terms of the
nucleon virtuality as follows
\begin{equation}
\delta_A(|{\bf p}|,E)= \left(1 - \frac{v_{NR}(|{\bf p}|,E)}{ 2\,
m_N\, \Delta E^{(N/A)}}\right)^{-2}, \label{allnuc}
\end{equation}
with  $v_{NR}(|{\bf p}|,E)$ given by Eq. (\ref{rem}). Note that
  using the
Koltun sum rule~\cite{koltun} corresponding to a Hamiltonian
containing only two-body forces, {\it viz}

\beq 2|\epsilon_A| = <E>-<T>\frac{A-2}{A-1}, \label{koltun0} \eeq
where $<T>$ and $<E>$ are the average kinetic and removal energies
per particle, respectively, one gets

\begin{eqnarray} &&
<v_{NR}>= -2m_N \left(\frac{A}{A-1}\frac{<{\bf p}^2>}{2m_N}
+<E>\right )= -4m_N\left[\left<T\right>+|\epsilon_A|\right],
\label{kolla}
\end{eqnarray}
so that the average value of Eq. (\ref{delgenerale})
(or Eq. (\ref{allnuc})) can  be written as follows

\ba
\langle
\delta_A(|{\bf p}|,E)\rangle  =
\left<
  \left  (1+
\frac
{E + \frac  {\mbox {$A$}}{\mbox{$A-1$}}   \frac{\mbox{${\bf p}^2$}}
    {\mbox{$2m_N$}}}
{\Delta E^{(N/A)}}
 \right )^{-2}
    \right> .
\label{deltamedio}
\ea

Since our derivation was non relativistic one cannot distinguish
the cases when  $v_{NR}$ or  $v$  are used. Hence, to check the
sensitivity to the higher order terms, we will also consider an
expression for $\delta_A$ in which $v$ is used instead of
$v_{NR}$:
\begin{equation}
\left<\delta_A(|{\bf p}|,E)\right>_v = \left<\left(1 -
\frac{v(|{\bf p}|,E)} { 2\, m_N\, \Delta
E^{(N/A)}}\right)^{-2}\right> ,\label{virtu}
\end{equation}
with  $v(|{\bf p}|,E)$ given by Eq. (\ref{rem}).

Eventually, we will also consider the partial
virtualities, i.e. the virtuality in a given state $f$, defined as
\begin{eqnarray} &&
<v_{NR}^{(f)}>= -2m_N \left(\frac{A}{A-1} <T>_f +<E>_f\right),
\label{kollapar}
\end{eqnarray}\\
and the corresponding partial coefficient of suppression of PLCs, i.e.
\ba\label{deltamediopar}
\langle
\delta_A^{(f)}(|{\bf p}|,E)\rangle  &=&
\left <
  \left  (1+
\frac
{E + \frac  {\mbox {$A$}}{\mbox{$A-1$}}   \frac{\mbox{${\bf p}^2$}}
     {\mbox{$2m_N$}}}
{\Delta E^{(N/A)}}
 \right )^{-2}
    \right >,
\label{effe}
\ea\\
satisfying

\begin{equation}
\left<\delta_A(|{\bf p}|,E)\right>= \sum_{f} \left<\delta_A^{(f)}
(|{\bf p}|,E)\right> ,\label{sum1}
\end{equation}

\noindent where the average  in Eq. (\ref{effe}) has to be taken  with
the proper partial Spectral Function.

If  we expand Eq. (\ref{deltamedio}),   we get its  lowest order
(LO) approximation in $({\bf p}/m_N)^2$ \ba \langle \delta_A(|{\bf
p}|,E)\rangle_{LO}  \approx \left( 1
-\frac{4\left(\left<T\right>+|\epsilon_A|\right)} {\Delta
E^{(N/A)}}\right).
 \label{approx} \ea

Let us now  address the physical reasons for the derived
 structure of Eq. (\ref{allnuc}).

It is often discussed in the literature that various properties of a
nucleon bound in
the nucleus should be modified due to the interactions with the
surrounding nucleons (usually referred to as  {\it medium
modifications}). Such a possibility  has been considered for the
case of the electromagnetic nucleon form factors, the parton
densities, and other quantities (see e.g.
\cite{Brown,FS85,Thomas1}). Medium  modifications are
theoretically often considered within the mean field approximation
and are assumed to depend on the mean
nuclear density,  with an implicit assumption that the
 modification does not depend on the momentum of the nucleon.
On the other hand,  we have just shown that  the contribution of
PLCs  exhibit a strong momentum dependence arising
 naturally from the reduction of the interaction  strength.
Accordingly,  one expects that in this model the modification of, e.g.
the  radius of a bound nucleon,   may  also depend upon the
nucleon momentum.
One intuitively expects that  possible modifications of the properties
of a bound nucleon should depend upon its off-shellness, which
can be expressed in terms of the nucleon virtuality as defined  by  Eq.(\ref{virt}).

To elucidate this point,  let us consider the electro-disintegration
of the deuteron, $e \,D \to e \,pn$,  as a function of the momentum of the
spectator nucleon $p_s$ (another option would be to consider  DIS  off the deuteron
in a tagged mode, i.e.  when the  spectator momentum is detected). The amplitude
$\mathcal A(\gamma^* + D \to pn)$ is an analytic
function of the Mandelstam variables, e.g.

\beq t=(p_D-p_s)^2 ,\eeq

\noindent i. e. the square of the momentum transfer and,
therefore, it  can be expressed as a series in terms of the
variable $m_N^2-t$. The  continuation to    the pole
 $t=m_N^2$ of the  propagator of the
interacting nucleon,  would correspond to  the interaction between
$\gamma^*$ and a free nucleon (this is analogous to the case of
the Chew-Low theorem relating the amplitude of the process $\pi +N
\to \pi \pi N$ to the $\pi -\pi$ scattering amplitude
\cite{Chew-Low}). Hence,  for small enough values of $m_N^2-t$,
the effect of medium  modifications are expected to  be
proportional to
 \beq
m_N^2-t =m_N^2-(p_d-p_s)^2=({\bf p^2}-m_N\,E_D)+ O({\bf p}^4,{\bf
p}^2\,E_D,E_D\,{\bf p}^2), \eeq which is exactly the functional
dependence of Eq. (\ref{deldeapp}). %%
Our reasoning  is heavily based upon the  analyticity of the
 amplitude in the  $t$ variable,
which justifies  the validity of the Taylor expansion near  the
 nucleon pole
in terms of powers of $(t-m_N^2)$.
Obviously our argument can be applied to the scattering off heavier
nuclei,  provided the residual (A-1) nucleon system has small enough
momentum  and excitation energy.  In this case   the relevant
perturbation parameter $\gamma_A$
is given  by Eq. (\ref{allnuc}), i.e.
\begin{equation}
\gamma_A(|{\bf p}|,E) = -(P_A-p_{s})^2+m_N^2\approx
2m\left(\frac{A}{A-1}\frac{{\bf p}^2}{2m_N}+E\right)= -v(|{\bf
p}|,E).
%=4m\left[\phantom{\frac12}\!\!\!\!\!\!\left<T\right>+|\epsilon_A|\right]
\label{gamma}
\end{equation}

%where the Koltun sum rule has again been used.
%%since we give as a function of p - no averaging is done - so
%%I removed the last line and reference to Koltum sr -
%since we discussed it already this is not a loss.
In the leading order of perturbation theory over binding effects
Eq. (\ref{deltamedio}) for the average value of $\delta_A$
 is recovered using the Koltun sum rule

\ba \langle \delta_A(|{\bf p}|,E)\rangle  = \left< \left( 1
+\frac{\gamma_A(|{\bf p}|,E)}{2m_N\Delta E^{(N/A)}}\right)
^{-2}\right> \approx \left( 1
-4\,\frac{\left<T\right>+|\epsilon_A|}{\Delta E^{(N/A)}} \right).
\label{deltamedio1} \ea

%**** Coefficient in the virial theorem needs to be checked *****
\subsection{PLCs and the variation of the  nucleon properties
in the medium}
 Denoting by  $\kappa(|{\bf p}|,E)-1$ the deviation from one of the ratio
of a certain  characteristic  (say the structure functions or the
radii) of the bound nucleon to its  vacuum value,  we can expect
that it will be given by \beq \kappa(|{\bf p}|,E) -1=
\frac{\gamma(|{\bf p}|,E)}{m_N^2}. \eeq

Therefore we can write
\ba
&&
\frac{\kappa(|{\bf p}|,E) - 1}{<\kappa(|{\bf p}|,E)> - 1}
={\gamma (|{\bf p}|,E)
\over
<\gamma (|{\bf p}|,E)>}
=\nonumber\\
&&={E +  \displaystyle\frac{A}{A-1}\displaystyle\frac{{\bf p}^2}
{2m}
\over
2\left[\phantom{\frac12}\!\!\!\!\!<T> + |\epsilon_A|\right]}
%\\&&\kappa(|{\bf p}|,E)
={2E\,m_N+{\bf p}^2{A\over A-1} \over
4m_N\left[\phantom{\frac12}\!\!\!\!\!\left<T\right>+|\epsilon_A|
\right]}, \ea
 which does not depend on the value
of $\Delta E$ or on the strength of
 the interaction for the probed
property. It follows,  from the above relation,
that the
region of small nucleon momenta and small excitation energies is the
least sensitive to
the effects of possible modifications of the nucleon
properties. Hence such a region is suitable  for the extraction of
the
properties of the free neutron from  scattering processes off the
deuteron and
$^3He$ using the analog of the Chew-Low procedure (see e.g. the discussion
in \cite{Jlab}). For the same reason,
the 3\% upper limit for  the change of the magnetic nucleon
radius  obtained from the analysis of the $Q^2$ dependence of
the inclusive
$(e,e')$ cross section near the quasi-elastic peak \cite{Sick},
 implies a much weaker
limit on the average change of the nucleon radius in nuclei. As a matter of fact,
in inclusive $(e,e')$ scattering, the cross section at the quasi
elastic peak ($x \simeq 1)$
is proportional to $\int d(|{\bf p}|) |{\bf p}| n_A(|{\bf p}|)$=
$<\frac{1}{|\bf p|}>$, giving
 $\left<{\bf p}^2\right > \simeq \la |{\bf p}|\ra\la 1/|{\bf p}|\ra$
whereas   in DIS
  ($x < 1$), it is directly proportional to  $\la |{\bf p}|^2\ra$; therefore, in quasi elastic
  scattering the average value of the probed  $\la |{\bf p}|^2\ra$
 is significantly smaller than the corresponding quantity in DIS,
  roughly by the factor

\beq
C=\displaystyle\frac{\la |{\bf p}|\ra}{
 \la |{\bf p}|^2\ra\la 1/|{\bf p}|\ra}.
 \label{factor}
 \eeq

As we have demonstrated,  the
study of the momentum dependence of the properties of a bound nucleon
 may be a better tool for the investigation of
modification effects  in nuclei. This can be achieved, e. g.,  by means
of semi-inclusive
processes (tagged
structure functions) and   by the measurement of the
momentum dependence of the ratio of electric to  magnetic
nucleon form factors.

\section{Suppression of PLCs in Inclusive Scattering. The A-dependence
 of the EMC effect}
Let us now illustrate how the  derived equations allow one to improve
 previous estimates of the A-dependence of the
  EMC effect in a particular region of $x$. To this
 end let us consider the
 well known EMC ratio for an isoscalar nucleus.
 \begin{equation}
 R_A(x,Q^2) \equiv  {AF_{2A}(x,Q^2)\over ZF_{2p}(x,Q^2) + N
 F_{2n}(x,Q^2)},
 \label{deno}
 \end{equation}
where $F_{2A}$ and $F_{2N}$ are the nuclear and nucleon structure
functions, respectively. We will considers in the Bjorken limit
two models for
 the
 the nuclear structure function $F_{2A}$, namely the light
 cone (LC)   and the virtual nucleon convolution (VNC) models. In both cases
 the the proton and neutron Spectral Functions and momentum distributions are
 considered to be the same.

 \subsection{The light cone  quantum mechanical model}

The light cone (LC)  quantum mechanics of nuclei
is based on the following assumptions: i) bound nucleons are on-shell
 ; ii) closure over final states is performed; iii) the light cone momentum of
 the nucleus is entirely carried by nucleons. Within this model the nuclear
 structure function $F_{2A}$ reads as follows
  \beq
F_{2A}^{LC}(x,Q^2) = A\int_x^A
\frac{d\alpha }{\alpha}d^2{\bf p}_\perp F_{2N}(x/\alpha,Q^2)
 \rho^{LC}(\alpha,{\bf p}_\perp),
\label{lightcone}
\eeq
where $\alpha =\frac{A}{M_A} p_{-}$ ($d^4\,p=\frac{M_A}{A} d\,\alpha
 d\,p_{+}d^2\,{\bf p}_{\perp}$ with $p_{\pm}$ as the
 corresponding light cone variables defined relative
 to the direction of the momentum transfer) is the light-cone fraction carried by the interacting nucleon
scaled to vary between zero and A,  and
 $\rho^{LC}(\alpha, {\bf p}_\perp)$ is the nucleon LC density matrix
 normalized according to  the baryon charge sum rule
\beq \int_0^A \frac{d\alpha }{\alpha}d^2{\bf p}_\perp
\rho^{LC}(\alpha,{\bf p}_\perp) = 1 ,
 \label{baryon} \eeq and
automatically satisfying the momentum sum rule \beq \int \alpha
\frac{d\alpha }{\alpha}d^2{\bf p}_\perp \rho^{LC}(\alpha,{\bf
p}_\perp) =1, \label{momentum} \eeq corresponding to $<\alpha>=1$.
In the non relativistic approximation for the nucleon motion
within a nucleus, LC  quantum mechanics coincides with the
conventional nuclear theory based on the non-relativistic
Schr\"odinger equation. An evident advantage of the LC mechanics
is the accurate account of relativistic effects,
%unnambigous identification of degrees of freedom,
 including those related to
pair production off vacuum resulting from the Lorentz
transformation.
% into parameters of effective Hamiltonian.

To calculate the effect of the suppression of PLCs in bound nucleons
 we have to substitute in the convolution formula $F_{2N}$ by $F_{2N}^{bound}=
 F_{2N}\,
\left<\delta_A(|{\bf p}|,E) \right>$, obtaining

\beq \frac{F_{2A}^{LC(\delta)}(x,Q^2)}{A} = \left<\delta_A(|{\bf
p}|,E) \right>\,\int_x^A \frac{d\alpha }{\alpha}d^2{\bf p}_\perp
F_{2N}(x/\alpha,Q^2)
 \rho^{LC}(\alpha,{\bf p}_\perp),
%<\delta_A(|{\bf p}|,E)>
\label{FALC}
\eeq
where $\left<\delta_A(|{\bf p}|,E) \right>$
is given by Eq.(\ref{delgenerale}) ( or  Eq.(\ref{allnuc})).
By expanding  $ F_{2N}(x/\alpha,Q^2)$ in Eq. (\ref{FALC})
%about $z_1
in a power series about  $\alpha=1$ one obtains
\begin{eqnarray}
\frac{F_{2A}^{LC}(x,Q^2)}{A} &\simeq& \left< \delta_A(|{\bf
p}|,E)\right> F_{2N}(x,Q^2) + x F_{2N}^{'}(x,Q^2)
<\,(\alpha-1)> +\nonumber\\
&+& \left[xF_{2N}^{'}(x,Q^2)\,+\,\frac{x^2}{2} F_{2N}^{''}(x,Q^2)\right]< (\alpha-1)^2>,
\label{expansion}
\end{eqnarray}
where the averages have to be evaluated with the light cone density
 $\rho^{LC}(\alpha, {\bf p}_{\perp})$.
  By considering that due to Eqs. (\ref{baryon}) and (\ref{momentum})
 $<(\alpha-1)> =0$ and taking the non relativistic limit
 to order $\frac{<{\bf p}^2>}{m_N^2}$  of the third term,
   the average values can be evaluated with the non-relativistic momentum distributions, obtaining
$\left<(\alpha-1)^2 \right>$=$\left< {\bf p}^2\right>/3m_N^2$ so
that
\ba
 \frac{F_{2A}^{LC(\delta)}(x,Q^2)}{A}\simeq \left<
\delta_A(|{\bf p}|,E)\right> F_{2N}(x,Q^2)
 + \left[ x F_{2N}^{'}(x,Q^2)
+\frac{x^2}{2} F_{2N}^{''}(x,Q^2)\right ]\frac{2<T>}{3m_N}.
\label{expNVC} \ea

 Supposing that the behaviour of $F_2(x)$ at
$x\le  0.5-0.7$ is governed solely by $u$-quarks ($u(x)\sim
(1-x)^n$ with $n\sim 3$)  one gets,

\ba
R_A^{LC(\delta)}(x,Q^2)=\frac{F_{2A}^{LC(\delta)}(x,Q^2)}{A\,F_{2N}
(x,Q^2)}\simeq \left<\delta_A(|{\bf p}|,E) \right> +
 nx \frac{x(n+1)-2}{6(1-x)^2}\,\,\frac{2<{T}>}{m_N}.
\label{expansionLC} \ea

\noindent yielding  the result of \cite{FS81NP} when $\delta_A=1$.
For $n=3$ we obtain
 \ba
R_A^{LC}(x,Q^2)=\frac{F_{2A}^{LC}(x,Q^2)}{A\,F_{2N}(x,Q^2)}\simeq 1
 +
 x\,\frac{(2x-1)}{(1-x)^2}\,\frac{2<T>}{m_N},
\label{expansionLC0} \ea leading to a cancelation of the Fermi
motion effects for $x = 1/2$. Hence $x$ $\sim 0.5$  is especially
convenient for the analysis, for one has, provided $\bf p$ is not
very large
 \ba R_A^{LC(\delta)}(x\sim 0.5)\propto\left<
\delta_A(|{\bf p}|,E)\right> \simeq \left<\left( 1 +\frac{
\gamma_A (p)}{2\,m_N\Delta
 E^{(N/A)}}\right)^{-2}\right> \simeq 1-4\frac{|\epsilon_A| + \left<T\right> }{\Delta
 E^{(N/A)}}.
\label{erreaFS} \ea\

 When $\delta_A=1$, one has, obviously,
\ba R_A^{LC}(x\sim 0.5) \simeq 1. \label{erreaFS0} \ea

We remind the reader that in order  to simplify the discussion, we
have up to now placed  $\lambda(x) = 0$ in the r.h.s. of Eq.
(\ref{deldeapp1}). Now, in order to compare our calculations with
the experimental data,  we necessitate an  explicit consideration
of the value $\lambda(x)$, so that we will use the following
expression for the ratio $R_A^{LC(\delta)}$
 \beq
 R_A^{LC(\delta)}(x\sim
0.5)\simeq 1-\left[ 1-\lambda(0.5) \right]4\frac{\left<T\right>
+|\epsilon_A|}{\Delta E^{(N/A)}}\,. \label{errelanda} \eeq

%Terms $\propto \epsilon_A $ are different but this effect is very small numerically%%

%***** need to check and discuss ******
%***********************\\
\subsection{The virtual  nucleon convolution model}
Another approach to the description of the DIS is the
approach in which   the role of the nuclear wave function is
played by the covariant vertex function described by appropriate
 Feynman diagrams.  In the case of DIS this approximation is usually refered to as
the virtual nucleon convolution model (VNC). In this approximation  the
interacting nucleon is  off-shell  and the light-cone fraction carried by the
interacting nucleon can be expressed through the
lab. frame momentum and energy of the residual system as

  \beq
  z= \frac{A}{M_A}(p_0- p_3),
  \label{alphavm}
  \eeq
  where
  \beq
  p_0 = M_A - \sqrt{(M_A-m_N+E)^2 + {\bf p}^2}.
  \eeq

  The nuclear structure function has the following form
\ba F_{2A}^{VNC}(x,Q^2)/A = \int F_{2N}(x/z,\,Q^2)f_A(z) d\,z
,\label{em} \ea
with the longitudinal momentum distributions $f_A^{N}$  given by\\
\ba f_A(z)= \int d^4\,p S_A(p)\,z\, \delta\left(z- \frac{A}{M_A}\
\left[p_0 - p_3 \right]\right).
\ea\\

The relation between the relativistic, $S_A(p)$),  and non
relativistic, $P_A(|{\bf p}|,E)$, Spectral Functions is,  to order
$(|{\bf p}|/m_N)^2$ ,  \beq S_A( p)= P_A(|{\bf p}|,E)\left[ 1+
{\mathcal O} (\frac{|{\bf p}|}{m_N})^2 + ...\right], \label{esseA}
\eeq
  and  baryon charge conservation is enforced by properly normalizing $f_A(z)$, i.e.
  \ba
\int f_A(z)d\,z= {C_A} \int d\,E\,d{\bf p}\, P_A(|{\bf
p}|,E)\,z\,d\,z \delta\left(z- \frac{A}{M_A}\left[p_o - |{\bf
p}|\cos\theta_{\widehat{\bf p\,q}} \right] \right)=1.
\label{baryon1}
\ea\\
The light-cone fraction carried by the nucleons in this model is
less than one \cite{FS87}
\begin{eqnarray}
<z >&=&\, \int z\,f_A(z) \,= 1\,-\,\frac{<E> -|\epsilon_A| +
<T_R>}{m_N}\equiv \eta <1, \label{convmom}
\end{eqnarray}
where $<T_R> \simeq \left< {\bf p}^2\right>/2(A-1)m_N$ and  $\eta$ is the total light cone
momentum carried by nucleons. The momentum sum rule is
restored by assuming that non nucleonic components carry the fraction
$1-\eta$ of the missing momentum.
These components should be added explicitly  to satisfy the momentum sum rule.
%%%what is stated below is not correct - one cannot both satisfy baryon and momentum sum rules in this formalism
% When only nucleonic degrees of freedom are considered,
%a common procedure is to renormalize the spectral Function so as  to satisfy the %momentum sum rule.
%Note that using realistic Spectral Functions, one has
%$\eta = 0.99,\, 0.99,\,
%0.98$
%in $^2H,\, ^3He,\,^4He$, respectively and $\eta\simeq 0.97$ for complex nuclei \cite{Simo}.
%%%it looks a bot strange that eta for D, and 3He is the same - I guess it is due to rounding the numbers.

By expanding  $ F_{2N}(x/z,\,Q^2)$ in Eq. (\ref{em}) about $z\sim
1$,  we obtain,  to order $\frac{<{\bf p}^2>}{m_N^2}$,

\begin{eqnarray}
F_{2A}^{VNC}(x,Q^2)/A &\simeq&  F_{2N}(x,Q^2) +
x F_{2N}^{'}(x,Q^2)
\frac{ < E > - |\epsilon_A|-\frac{2}{3}<T> + <T_R>}{m_N} +\nonumber\\
&+& \left[xF_{2N}^{'}(x,Q^2)\,+\,\frac{x^2}{2}
F_{2N}^{''}(x,Q^2)\right]\frac{2<T>}{3m_N}. \label{expNVC1}
\end{eqnarray}
%which differs from the LC model expansion (Eq. (\ref{expNVC})) by the presence of the second term
%in the $rhs$ which is proportional  to $(1-\eta)$, i.e. the  momentum carried by non-nucleonic
% degrees of freedom, and resulting from the violation of the momentum sum rule (cf. Eq.
% (\ref{convmom})).
%Supposing again
%that the behavior of $F_{2N}(x)$ at $x\le  0.5-0.7$ is governed solely
%by $u$-quarks ($u(x)\sim (1-x)^3$) and using the Koltun sum rule (Eq. (\ref{koltun0})),   one gets to  order $\frac{<{\bf p}^2>}{m_N^2}$
%\begin{eqnarray}
%&&R_A^{VNC(\delta)}(x,Q^2)=\frac{F_{2A}^{VNC(\delta)}(x,Q^2)}{A\,F_{2N}(x,Q^2)}
%\simeq \nonumber\\
%&&  \simeq \left< \delta_A(|{\bf p}|,E) \right> - \frac{3x}{(1-x)}\frac{
% |\epsilon_A|+\frac{1}{3}<T> }{m_N}\,
%+ \,x\,\frac{(2x-1)}{(1-x)^2}\,\frac{2<T>}{m_N}
%\label{expNVC2}
%\end{eqnarray}
%and
%\begin{eqnarray}
%&&R_A^{VNC}(x,Q^2)=\frac{F_{2A}^{VNC}(x,Q^2)}{A\,F_{2N}(x,Q^2)}\simeq \nonumber\\
%&&\simeq 1
%-\frac{3x}{(1-x)}\frac
%{ |\epsilon_A|+ \frac{1}{3}<T> }{m_N}+
% x\,\frac{(2x-1)}{(1-x)^2}\,\frac{<T>}{m_N}
%\label{expNVC0}
%\end{eqnarray}
%when $\delta_A=1$.
Choosing again  $x=0.5$ we obtain,
\begin{eqnarray}
R_A^{VNC}(x\sim 0.5) =1-3\,\frac{ |\epsilon_A| +
\frac{1}{3}<T>}{m_N}. \label{erreaVNC0}
\end{eqnarray}

We notice that A-dependence of all terms contributing to ${R}$ is
very similar since all coefficients are dominated by the
contribution of the kinetic energy term. Hence, independent of the
details one expects an approximate factorization:

\beq R_A(x,Q^2)-1=\frac{\phi(x,Q^2)}{f(A)}, \label{ratio}
 \eeq which works
well experimentally.

An important quantity considered in the literature is the relation
between the  EMC effect ratio  in the deuteron $R_D(x,Q^2)$, given
by Eq.(\ref{deno}), and the value of $R_A(x,Q^2)/R_D(x,Q^2)$
measured experimentally.
 Such a relation has been used to extract the
neutron to proton ratio  $F_{2n}/F_{2p}$. Previous
estimates gave for $^{56}Fe$ \cite{FS85,FS88}
\begin{equation}
 R_D(x,Q^2)-1 = c\,\left(\frac{R_A(x,Q^2)}{R_D(x,Q^2)}\,-1\right)\qquad c\,=
 \,\frac{1}{4}.
 \label{adeut}
\end{equation}\\
 This relation is referred to by Bodek \cite{Bodek} as a {\it density model}, since
in the mean field approximation  the average kinetic energy is proportional to the average nuclear density
(note however that the average nuclear density is hardly defined for light nuclei while
expressions  (\ref{erreaFS}) and  (\ref{erreaVNC0}) are   well defined even for A=2).
It is also worth emphasizing that $<{\bf p}^4>/<{\bf p}^2>$  is  significantly smaller
 in the deuteron than in  heavy
nuclei,  so that as soon as  the terms proportional to $<{\bf
p}^4> $ become important, Eq. (\ref{adeut}) breaks down, which
occurs at to $x\sim 0.7-0.8$. At  $x=1$  Eq. (\ref{adeut}) is
badly violated, since the r.h.s.  remains finite in this limit,
while the l.h.s. tends to infinity.

Relation (\ref{adeut})  has been  used in several papers (see e.g.
\cite{Bodek}) to extract the neutron to proton ratio
$F_{2n}/F_{2p}$. We will show in the next Section that by using
realistic nuclear spectral functions a different relation will be
obtained.

We will explore in the next section the sensitivity of the
A-dependence of the EMC effect predicted by our models.

\section{Results of calculations}

In this Section  the results of of our calculations based upon
realistic Spectral Functions for few-nucleon systems and complex
nuclei are presented. We have  calculated the following
quantities:
\begin{enumerate}
\item the normalization ($S$) and the average values of the kinetic ($<T>$) and removal
($<E>$) energies in $^3He$  corresponding to various states of the spectator two-nucleon
system. The results are presented in Table I.
\item The same quantities as in Table I are reported in Table II for $A=2$ and
$4 \leq A \leq 208$.

\item Various average values of powers  of the nucleon momentum ${\bf p}$, which are listed in
 Table III.

\item The average values (divided by $2\,m_N$)  of the virtuality in the states
 $f$ $<v_{NR}^{(f)}>$  (Eq. (\ref{kollapar}))  and their sum  (Eq. (\ref{kolla})). The results are
listed in Table IV.

\item The average value of the coefficient $\delta ({\bf p}, E)$
of the suppression of PLCs in various configurations
(Eq.(\ref{effe}))
and their sum (Eq.(\ref{sum1})), together with $\delta^{(v)}$ (Eq.(\ref{virtu})),
 presented in Table V. In the case of $^3He$
 $\la\delta^{(0)}\ra
\equiv \la\delta^{gr}\ra$ and  $\la\delta^{(1)}\ra
\equiv \la\delta^{ex}\ra$ (cf. the sentences after  Eq. (\ref{delge3})).

\item The EMC ratio
 given   by
 \beq
 R_A(x,Q^2)=\frac{F_{2A}(x,Q^2)}{F_{2D}
(x,Q^2)}, \label{true}
 \eeq
  calculated at  $x=0.5$ with Eqns. (\ref{erreaFS}), (\ref{erreaFS0})
 and (\ref{erreaVNC0}) (all multiplied by $AF_{2N}/F_{2D}$, to be consistent with the experimentally measured
$R_A(x,Q^2)$; note, in fact, that Eq. (\ref{true})) differs from
  Eq. (\ref{deno}) in that the denominator represents the deuteron structure function and
  not the sum
  of the nucleon structure functions
  $F_{2N}$) and compared with SLAC experimental data
 fitted by
  ${R}^{exp}=1.009\,A^{-0.0234}$ \cite{Gomez} .

   In the case of the LC with
  suppression of PLC's we have used   both  $\Delta E^{(N/D)}=800 MeV$ and
$\Delta E^{(N/A)}\sim  500 MeV$ and $\Delta E^{(N/D)}=
\Delta E^{(N/A)} = 800 MeV$.  Wave functions and Spectral functions
as in Tables \ref{tab1} and \ref{tab2}.

\end{enumerate}

In our calculations we have  employed the spectral function of
$^3He$ given  in Ref. \cite{leonya3} obtained using the
ground-state three-body wave function from the Pisa Group
\cite{pisa}  corresponding to the $AV18$ interaction  \cite{av18}.
For complex nuclei, calculations have been  performed with the
model spectral function of Ref. \cite{CS}, which correctly
reproduces the momentum and energy distributions as obtained from
realistic calculations on complex nuclei.

The following comments  concerning the obtained results are in order:

\begin{enumerate}
\item {\it Tables I and II.}
The average kinetic and removal energies in channels $ex(f)$ are much larger than the corresponding
quantities in channels $gr(0)$ and the high momentum components are linked to high removal energies,
which is a well known result demonstrated long ago \cite{CPS}.

\item  {\it Table III.}

The effects of correlations on the  high momentum components is clearly seen. The value of the
quantity $C$ (Eq. (\ref{factor})) indicates that the probed value of   $<{\bf p}^2>$
in Deep Inelastic Scattering  is larger by a factor $2$ than in  Quasi Elastic Scattering.

\item  {\it Tables IV and  V.}

The nucleon virtuality in states $ex(f)$ is much higher than in states $gr(0)$ due to the
higher average values of the removal energy and momentum.
Consequently, the suppression
of PLCs  in states $ex(f)$ is expected to be  higher than in states $gr(0)$.
Semi-inclusive processes with the spectator $A - 1$ nucleus in
high excited states should  be a very effective tool to investigate  the suppression of PLCs.

\item {\it Table VI.}

Both the LC model with suppression of PLC's  and the VNC model predict almost no A-dependence
of the EMC effect in the range $4 \leq A \leq 56$. The results pertaining to the former model do depend upon
the value of $\Delta E^{(N/D)}$. We have tried both  $\Delta E^{(N/D)}=800 MeV$ and
$\Delta E^{(N/A)}\sim  500 MeV$ and $\Delta E^{(N/D)}=
\Delta E^{(N/A)} = 800 MeV$; in the former case  to reproduce the magnitude of the EMC effect
 for iron at
$x=0.5$ one needs $\lambda \sim 0.4$, which is similar to the
value used  in Ref. \cite{FS88}.
\end{enumerate}

\begin{figure}[h] %%%%         Fig.1
%\epsfxsize 5.2in
%\centerline{ \epsfbox{fig2.eps}}
\begin{center}
\includegraphics[height=10cm,width=14cm]{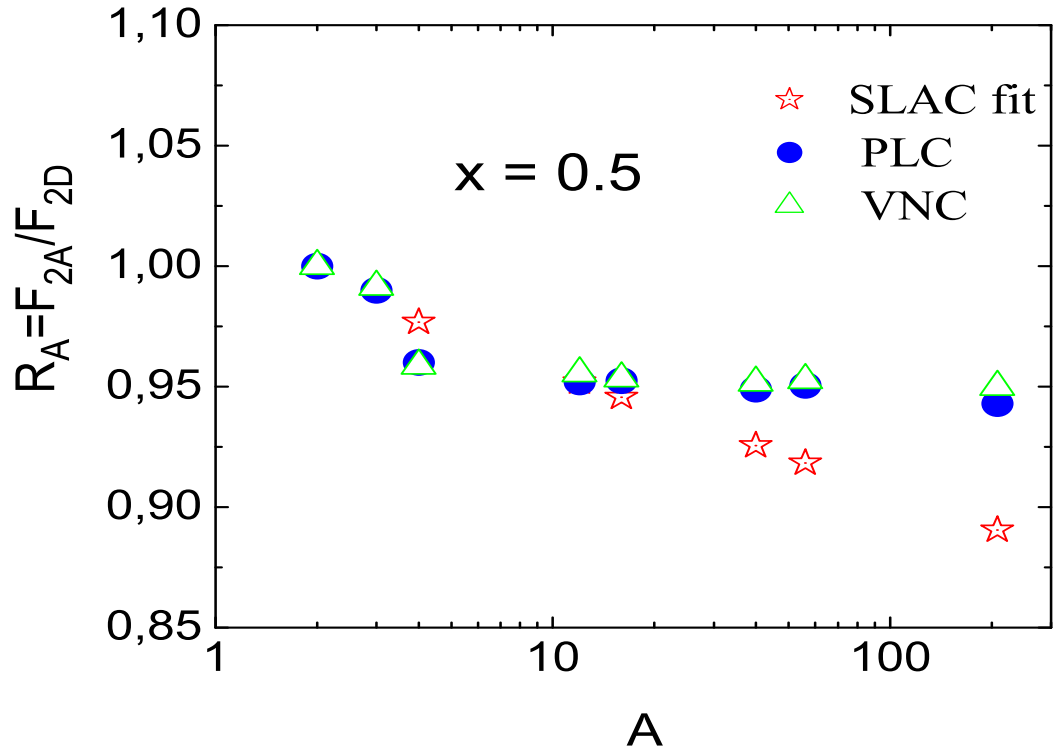}
\vskip -0.5cm \caption{ The EMC ratio $F_{2A}/F_{2D}$ (Eq.
\ref{true}) at $x=0.5$ corresponding to the values given in Table
VI. Note that the  SLAC  fit \cite{Gomez} to the experimental data
  ${R}^{exp}=1.009\,A^{-0.0234}$  does not  include
systematic and statistic errors
 and   has a tendency to underestimate the effect for $^4He$.
 }
  \label{Fig1}
  \end{center}
\end{figure}

In order to better illustrate the A-dependence of the EMC effect, we show in Fig. \ref{Fig1}
 the results presented in Table VI normalized   to the SLAC experimental value of ${R}$ for carbon
%(here we define ${R}$  using in the denominator the free nucleon structure function).
It can be clearly seen that, at variance with the trend of the
SLAC data, the VNC model does not predict, in the interval $4 \le
A \le 208$, any A-dependence. A flattening of the A-dependence of
$R$ is also predicted by the LC approach with suppression of PLC's
which appears to be sensitive to
  the value
   of $\Delta E^{(N/A)}$. Note that  our results seem to agree
   with recent experimental data from JLab
   \cite{newJlab} which  exhibit practically the same EMC effect for $^4He$
   and $^{12}C$. It should also be pointed out, in this respect,  that for heavy nuclei
   ($A \ge 50$) the Coulomb effect
   which we will discuss below, leads to an additional, and increasing with A,
   suppression of the EMC ratio.

Our studies of the A-dependence of the EMC effect allow us to make
predictions for the EMC effect in the deuteron which will be
useful for a comparison  with the forthcoming JLab data aiming  at
determining $F_{2n}(x,Q^2)$ from the measurements of the deuteron
tagged structure function. Our results  suggest that
 \beq
 R_D(x,Q^2) -1=c_A(x) \left( \frac{R_A}{R_D} -1 \right),\,\
  \label{new}
 \eeq
 with $c_A(x)$ practically independent
  of $x$ for $0.3 \le x \le 0.6$. Within the Virtual Nucleon
  Convolution  model we obtain   $c_{12}(x) \simeq 0.34$ and
  $c_{ 56  }(x)= 0.33$, whereas in the LC model with suppression
  of PLCs we have
  $c_{12}(x) \simeq 0.23$ and
  $c_{ 56  }(x)= 0.22$, in the case of  $\Delta E^{N/A}\neq \Delta E^{N/D}$,
 and $c_{12}(x)\simeq 0.43$  and  $c_{56}(x) = 0.41$, in the case
 of
  $\Delta E^{N/A}=\Delta E_{N/D}= 800 MeV$. Our results, which  are  somewhat different from the estimate
  $c_{56}(x)=1/4$, might   have consequences on the extraction of the ratio
  $F_{2n}/F_{2p}$.

\section{Heavy nuclei and the Coulomb field effect.}

The reasons for a possible   increase of the EMC effect in heavy
nuclei  like, e.g. higher order effects in the nuclear density
which are not included in the current treatment, for example the
effects due to three body forces. Whereas such an effect might
also  be present in medium weight and light nuclei, an effect
which is specific only  for heavy nuclei is the presence of the
coherent, Weizecker-Williams (WW) photon field of the nucleus
which is due to the fact that the fields of individual protons add
coherently, leading to a larger momentum fraction carried by the
photon field in nuclei as compared to that carried by  individual
free protons. Here we will calculate this effect in the leading
order in $Z$. The WW expression for the spectrum of the equivalent
photons emitted in the case of a finite size nucleus is given by

\begin{equation}
F_A^{\gamma}(x)={\alpha_{em}\over \pi x}
{Z^2\over A}
\int  dk_t^2 \frac{F_A^2(k_t^2+x^2 m_N^2)}{(k_t^2+x^2 m_N^2)^2},
\label{ww1}
\end{equation}\\
\noindent  where $k_t$ is the transverse momentum of the photon,
and the finite  sizes of the nucleus  enter via the presence of
the nuclear e.m. form factor
 $F_A(t)$ which we take in the exponential form, i. e.
 $F=\exp(-R_A^2 (k_t^2+x^2m_N^2)/6)$. This allows one  to perform the
 integral over the transverse momenta of the photons obtaining
 \begin{equation}
F_A^{\gamma}(x)=
  \frac{2 \alpha_{em}Z^2}{Ax\pi} \ln \left(\frac{\sqrt{3}}{R_Am_N x}\right) \exp (-R_A^2m_N^2x^2/3).
\end{equation}\\
 From this one can find  the contribution of the photons to the momentum sum rule
 in the following form
 \begin{equation}
 {2\alpha_{em}\over \pi}  {Z^2\over A}  \int\limits_0^1 dx \ln (\sqrt{3}/x m_NR_A)
\exp(-R_A^2m_N^2 x^2/3)\approx
 {2\alpha_{em}\over \pi}  {Z^2\over A}
 {\sqrt{3}\over m_NR_A}.
 \label{ww5}
 \end{equation}\\
%%here integral over x is taken approximately
In the case of a lead target this leads to a  fraction carried by
photons of about $ 1\%$ (we give here the number obtained by
numerical evaluation of the integral in Eq. (\ref{ww5})).

 The suggestion that photons can carry non-negligible  fraction of the nucleus
momentum, helping therefore the explanation of the EMC effect in
heavy nuclei has been previously made in Ref.
\cite{Birbrair:1989zf}. Our calculation was done directly in the
infinite momentum frame whereas in Ref. \cite{Birbrair:1989zf}
virial  theorem has been used. The numerical results are however
quite similar.

It should be  moreover be mentioned  that the photon momentum is
taken from the fraction of the momentum carried by protons.
Accordingly, in a heavy nucleus protons  loose a fraction of
momentum of $\sim 2.5\%$, leading to an enhancement of the
contribution of this effect to the EMC effect. Note here that a
loss of $5\%$ fraction of the momentum carried by nucleons in
nuclei (corresponding to the quantity  $1- \eta$, $\eta$ being
defined by  Eq. (\ref{convmom})) would be sufficient to explain
the whole EMC effect for heavy nuclei. Since $F_{2p}/F_{2n} \ge 2$
at $x \ge0.5$, the discussed effect can explain approximately
$1/3$ of the overall EMC ratio in heavy nuclei. This point will be
discussed in more detail elsewhere.

\section{Conclusions}

We have provided a derivation of the suppression of the
PLC's  of  a nucleon bound in the nucleus $A$
 with the spectator nucleus $A-1$  being in a particular energy configuration.
 We have pointed out that  that the  result we have obtained can be interpreted
  as a specific dependence of the nucleon deformation upon  the nucleon virtuality and argued
   that  such a pattern is of quite general nature for small  values of
    excitation energy and the momentum of the   nucleon.
   Within such a framework, we have discussed the effects of the nucleon virtuality
   on the  investigation of the modification of the radius and  form factor of nucleons
    embedded in the
   nuclear medium, illustrating that Deep Inelastic Processes  might be more effective than
   Quasi-elastic processes,  in that in the former   higher momentum component are probed.
    Eventually,  we have illustrated the
implications of our approach for the A-dependence of the EMC
effect, obtaining very similar effects for $^4He$ and $^{12}C$
which agree well with the preliminary JLab data \cite{newJlab}. We
also suggested that a substantial part of the increase of the EMC
effect for heavy nuclei could be due to the Coulomb  field
effects. %%

\acknowledgments{ We would like to thank A. Umnikov for his
contribution in the early stage  of this study. One of us (MS)
thanks A. Thomas for a discussion on the nucleon deformation in
the mean field approximation. This work is supported in part by
the U.S. Department of Energy and by the Italian Ministry of
Research and University, through funds PRIN 05. L. Frankfurt
thanks support from a German-Israel Grant (GIF) and L. P. Kaptari
would like to thank  of INFN, Sezione di Perugia and Department of
Physics, University of Perugia for support and hospitality. }
\newpage
\appendix
\section{The non relativistic spectral functions of ${\bf ^3He}$}
\label{ap:1}

For $^3He$ one has to define the $proton$ and $neutron$ spectral functions formed by the three different channels $^3He \to n(pp)$, $^3He \to p(D)$, and $^3He \to p(np)$, viz.
\begin{eqnarray}
&&P_p(|{\bf  p}|, E) = P_{p(D)}(|{\bf  p}|, E)+P_{p(pn)}(|{\bf  p}|, E),
\label{Sp-he}\\
&&P_n(|{\bf  p}|, E) = P_{n(pp)}(|{\bf  p}|, E),
\label{Sn-he}
\end{eqnarray}
where $ P_{p(D)}$ is usually referred to as the "ground"
spectral function,  $P_{p}^{gr}$, and $P_{p(pn)}$ and $P_{n(pp)}$
  the "excited" spectral functions,
$P_{p}^{ex}$ and $P_{n}^{ex}$, respectively~\cite{CPS}.
This terminology
 refers to
the final spectator state or, in other words, to the state
of the residual $A-1$-system.
The spectral  functions are normalized in the following way:
\begin{eqnarray}
&&\int P_p(|{\bf  p}|, E)dE |{\bf  p}|^2d|{\bf  p}| = 1,
\quad
\int P_n(|{\bf  p}|, E)dE |{\bf  p}|^2 d|{\bf  p}| = 1,
\label{Snorm-he}
\end{eqnarray}
therefore the isotopic factors and angular factors have to be explicitly
 taken into account.
The mean values of the kinetic and separation energies are then given by:
\begin{eqnarray}
&&\langle E_p \rangle = \int P_p(|{\bf   p}|, E)E dE |{\bf  p}|^2d|{\bf  p}|,
\quad\quad
\langle E_n \rangle = \int P_n(|{\bf  p}|, E)E dE |{\bf  p}|^2 d|{\bf  p}|,
\label{SE-he}\\
&&\langle T_p \rangle=\int P_p(|{\bf   p}|, E)\frac{|{\bf  p}|^2}{2m}
 dE |{\bf  p}|^2d|{\bf  p}|
,
\quad
\langle T_n \rangle=\int P_n(|{\bf  p}|, E)\frac{|{\bf  p}|^2}{2m}
 dE |{\bf  p}|^2 d|{\bf  p}|.
\label{ST-he}
\end{eqnarray}
and the corresponding averages for the nucleon are
\begin{eqnarray}
 && \langle E_N \rangle = \frac{2}{3}\langle E_p \rangle +\frac{1}{3}
 \langle E_n \rangle,
\label{SEN-he}\\
&& \langle T_N \rangle = \frac{2}{3}\langle T_p \rangle +\frac{1}{3}
\langle T_n \rangle.
\label{STN-he}
\end{eqnarray}

Using the Koltun sum rule~\cite{koltun}
\begin{equation}
\langle E \rangle  = 2|{\epsilon_A}|+ {\frac{A-2}{A-1}}{\langle T
\rangle }
\label{koltun1}
\end{equation}
  the effective
binding energies per nucleon for all components can be defined:
The corresponding averages for the nucleon are:
\begin{eqnarray}
&& \epsilon_{N} = \frac{1}{2}
\left (\langle E_N \rangle - \frac{1}{2}\langle T_N \rangle \right ),
\label{epsN-he}\\
&& \epsilon_{p} = \frac{1}{2}
\left (\langle E_p \rangle - \frac{1}{2}\langle T_p \rangle \right ),
 \label{epsp-he}\\
&& \epsilon_{n} = \frac{1}{2}
\left (\langle E_n \rangle - \frac{1}{2}\langle T_n \rangle \right ),
\label{epsn-he}\\
&& \epsilon_{N} = \frac{2}{3}\epsilon_{p} + \frac{1}{3}\epsilon_{n},
\end{eqnarray}
with the binding energy per particle of $^3He$, $\epsilon_3$  given by
 $3\times \epsilon_{N}$.

The mean values associated to   $P^{gr}$ and $P^{ex}$, are given
by eqs. (\ref{Sp-he})-(\ref{Sn-he})(\cite{CS}):
\begin{eqnarray}
&& \langle P^{gr}\rangle = \int P^{gr}(|{\bf  p}|, E)dE |{\bf  p}|
^2d|{\bf  p}|, \quad\quad \langle P^{ex}\rangle = \int
P^{ex}(|{\bf p}|, E)dE |{\bf  p}|^2d|{\bf  p}| ,
\label{Sgrex}\\
&&\langle E^{gr}\rangle = \int P^{gr}(|{\bf  p}|, E)EdE |{\bf
p}|^2d|{\bf  p}| , \quad
 \langle E^{ex}\rangle = \int P^{ex}(|{\bf  p}|, E)EdE |{\bf  p}|^2d|{\bf  p}| ,
\label{Egrex}\\
&& \langle T^{gr}\rangle = \int P^{gr}(|{\bf    p}|, E)\frac{|{\bf
p}|^2}{2m}dE |{\bf  p}|^2d|{\bf  p}| ,\, \langle T^{ex}\rangle =
\int P^{ex}(|{\bf  p}|, E)\frac{|{\bf p}|^2}{2m}dE |{\bf
p}|^2d|{\bf  p}| ,
\label{Tgrex}\\
&& \quad \epsilon^{gr} = \frac{1}{2} \left (\langle E^{gr} \rangle
- \frac{1}{2}\langle T^{gr} \rangle \right ), \quad\quad\quad\quad
 \epsilon^{ex} = \frac{1}{2}
\left (\langle E^{ex} \rangle - \frac{1}{2}\langle T^{ex} \rangle \right ).
\label{epsgrex}
\end{eqnarray}

The values of the various quantities which will be used in this paper  are listed in Table ~\ref{tab1}.
They have been obtained with
the spectral function of Ref. \cite{leonya3} and correspond to the Wave function of the Pisa
Group \cite{pisa} obtained variationally using the AV18 interaction \cite{av18}.

\newpage
%

%%%%%%%%%%%
\newpage
%%%%%%%%%%%%%%%
\begin{table}[h]
\caption{The normalization  factors $S_f$ (Eq. (\ref{norm})). the mean kinetic, $<T>$ , and removal, $<E>$,
energies, and the energy per nucleon $|\epsilon_3|$ for Helium-3, calculated with the spectral function of Ref.
\cite{leonya3} and the Pisa group wave function \cite{pisa} corresponding to the $AV18$ interaction \cite{av18}.
 The state $f=gr$ corresponds to the
spectator proton-neutron system in the ground state (a deuteron), whereas the state
$f=ex$ corresponds to the proton- neutron or proton-proton systems in the continuum}
\begin{tabular}{ccccccccccccc}\hline\hline
 & \multicolumn{3}{c}{ \,\,\, Norm, S\,\,\, } &\multicolumn{3}{c}{$ \,\,\,\langle T \rangle$, MeV\,\,\,}
 &\multicolumn{3}{c}{$ \,\,\,\langle E \rangle$, MeV\,\,\,}
 &\multicolumn{3}{c}{$ \,\,\,|\epsilon_3|$, MeV\,\,\,}\\
\hline
   &gr \,\,\,&\,\,\,ex &\,\,\,tot.\tablenotemark[1]\,\,\,\,\,\,
   &gr\,\,\, &\,\,\,ex &\,\,\,tot.\tablenotemark[1]\,\,\,\,\,\,
   &gr\,\,\, &\,\,\, ex&\,\,\, tot.\tablenotemark[1]\,\,\,\,\,\,
   &gr\,\,\, &ex\,\,\, &\,\,\,tot.\tablenotemark[1]\,\,\,\,\,\, \\
\hline
proton  &0.65 &0.35  &1 &4.67 &8.60  &13.27  &3.72 & 6.81    & 10.53    &0.69 &1.26 & 1.95\\
neutron &    0&   1  &1 &0    &17.69 &17.69  &0    & 16.33   & 16.33    &0    &3.74 & 3.74\\
per nucleon \tablenotemark[2]  &-    &-     &1 &3.11 &11.63 &14.74  &2.48  & 9.99  & 12.47  &0.46  &2.09 & 2.55
\tablenotemark[3]\\ \hline\hline
\end{tabular}
\tablenotetext[1]{total = gr+ex}
\tablenotetext[2]{ per nucleon = (2 proton+neutron)/3}
\tablenotetext[3]{ $3\times \epsilon_3= E_3 \approx 7.7$ MeV,
the value computed in refs.~\protect\cite{pisa}}
\label{tab1}
\end{table}
\newpage
\begin{table}[h]
\caption{The same as in Table \ref{tab1} but for the deuteron and complex nuclei.
The results for $A=2$ correspond to the $AV18$ interaction and the ones for
$4\leq A \leq 208$ to the Spectral Function of  Ref. \cite{CS}.}
\begin{tabular}{|c|c|c|c|c|c|c|c|c|c|}\hline\hline
$A$ & $S_0$ & $S_1$ & $<T>_0$ & $<T>_1$ & $<T>$ & $<E>_0$ & $<E>_1$ & $<E>$ & $|\epsilon_A|$\\
&&&$MeV$&$MeV$&$MeV$&$MeV$&$MeV$&$MeV$&$MeV$\\
\hline %\tablenotemark[1]
$D$       &1.0 & -  & -    &      &11.07 &      &      & 2.226 &1.113\\
$^4He  $  &0.8 &0.2 & 8.23 &17.55 &25.78 &15.85 &19.20 &35.05 &8.93\\
$^{12}C$  &0.8 &0.2 &13.54 &18.93 &32.47 &18.40 &26.55 &44.95 &7.72\\
$^{16}O$  &0.8 &0.2 &11.22 &19.73 &30.95 &19.42 &27.20 &46.62 &8.87\\
$^{40}Ca$ &0.8 &0.2 &13.39 &20.45 &33.84 &21.28 &28.57 &49.85 &8.44\\
$^{56}Fe$ &0.8 &0.2 &11.45 &21.26 &32.71 &20.00 &29.06 &49.06 &8.47\\
$^{208}Pb$&0.8&0.2  &14.72 &24.40 &39.12 &18.53 &34.79 &53.32 &7.19\\
 \hline\hline
\end{tabular}
%\tablenotetext[1]{total = gr+ex}
\label{tab2}
\end{table}
\newpage
\begin{table}[h]
\caption{Various average values  of the nucleon momentum ${\bf p}$. Wave functions and Spectral Functions
as in Tables \ref{tab1} and \ref{tab2}.}
\hspace*{-1cm}
\begin{tabular}{|c|c|c|c|c|c|c|c|c|c|c|}\hline\hline
$A$ & $<{\bf p}^2>_0$ & $<{\bf p}^2>_1$ & $<{\bf p}^2>$ & $<|{\bf p}|>_0$ &
   $<|{\bf p}|>_1$ & $<|{\bf p}|>$ & ${<|{\bf p}|>_0}^{-1}$ & ${<|{\bf p}|>_1}^{-1}$ &
   ${<|{\bf p}|>}^{-1}$ & $\frac{<|{\bf p}|>}{<{\bf p}^2>\,<1/<|{\bf p}|>}$\\
&$fm^{-2}$&$fm^{-2}$&$fm^{-2}$&$fm^{-1}$&$fm^{-1}$&$fm^{-1}$&$fm$&$fm$&$fm$&\\\hline
$D$       &0.533 & -         &0.533&0.502 & -&0.502 &3.74 & -&3.74 &0.25\\
$^3He$    & 0.150&0.5651     &0.71 &0.224 &0.415 &0.64  &1.31 &0.548 &1.86 &0.49\\
$^4He$    & 0.396&0.845      &1.24 &0.517 & 0.346&0.86  &1.59 &0.19 &1.78 &0.39\\
$^{12}C$  & 0.652& 0.912     &1.56 & 0.6777&0.369 &1.05  &1.13 &0.17 &1.30 &0.51\\
$^{16}O$  &0.540 &0.951      &1.49 &0.618 &0.380 &0.998 &1.24 &0.157 &1.40 &0.48\\
$^{40}Ca$ &0.645 & 0.985    &1.63 & 0.679&0.385 &1.06  &1.13 &0.16 &1.29 &0.51\\
$^{56}Fe$ &0.552 &1.024     &1.58 & 0.631&0.400 &1.03  &1.20 &0.14 &1.34 &0.49\\
$^{208}Pb$&0.709 &1.176     &1.88 & 0.719&0.480 &1.20  &1.03 &0.18 &1.21 &0.53\\
 \hline\hline\end{tabular}\label{tab25}\end{table}
\newpage
\begin{table}[h]
\caption{The quantities  $<v_{NR}^{(f)}>/2 m_N$ where  $<v_{NR}^{(f)}>$
is the average value of the virtuality in various states
 (Eq. (\ref{kollapar})
) and their sum  $<v_{NR}>/2m_N =-2(<T> + |\epsilon_A|)$ (cf. the discussion after  Eq. (\ref{kolla})).
 In the case of $^3He$,
 $\la v_{NR}^{(0)}\ra
\equiv \la v_{NR}^{(gr)}\ra$ and  $\la v_{NR}^{(1)}\ra
\equiv \la v_{NR}^{(ex)}\ra$ (cf. Eq. (\ref{delge3})).  Wave functions and Spectral Functions
as in Tables \ref{tab1} and \ref{tab2}.  All quantities  in MeV.}
\begin{tabular}{|c|c|c|c|}\hline\hline
 A  & $\la v_{NR}^{(0)}\ra$ &
  $\la v_{NR}^{(1)}\ra$ & $\la v_{NR}\ra$\\\hline
$^3He  $   &  -7.15 & -27.44 & -34.59\\
$^4He  $   & -26.82 & -42.58 & -69.40\\
$^{12}C$   & -33.17 & -49.11 & -82.28\\
$^{16}O$   & -31.40 & -48.28 & -79.68\\
$^{40}Ca$  & -35.00 & -49.54 & -84.54\\
$^{56}Fe$  & -31.66 & -50.76 & -82.44\\
$^{208}Pb$ & -32.87 & -59.33 & -92.20\\
 \hline\hline
\end{tabular}
\label{tab31}
\end{table}
\newpage
\begin{table}[h]
\caption{The average value of the coefficient of the suppression of PLCs in various
configurations (Eq.(\ref{effe}))
and their sum (Eq.(\ref{sum1})). The latter is compared with the results of  Eq.(\ref{virtu}),
 where the nonrelativistic
reduction  of the virtuality (Eq.(\ref{rem})) has not been performed. The last
column exhibits the lowest order value
(in $({\bf p}/m_N)^2)$   given by Eq. (\ref{approx}) represents
the results of Eq. (\ref{approx}).
 In the case of $^3He$
 $\la\delta^{(0)}\ra
\equiv \la\delta^{(gr)}\ra$ and  $\la\delta^{(1)}\ra
\equiv \la\delta^{(ex)}\ra$ (cf. the discussion after Eq. (\ref{delge3})).
  Wave functions and Spectral Functions
as in Tables \ref{tab1} and \ref{tab2}.}
\begin{tabular}{|c|c|c|c|c|c|}\hline\hline
 A  & $\la\delta^{(0)}(|{\bf p}|,E)\ra$&
  $\la\delta^{(1)}(|{\bf p}|,E)\ra$ &$\la\delta(|{\bf p}|,E)
  \ra$ &$\la\delta (|{\bf p}|,E)\ra_v$ & $\la\delta (|{\bf p}|,E)\ra_{LO}$ \\\hline
$D$        & 0.95 & -  &0.95 & 0.96   & 0.94                                      \\
$^3He  $   & 0.43 &0.48  &0.91 & 0.92 & 0.90                                    \\
$^4He  $   & 0.70 & 0.12 &0.82 & 0.82 & 0.72                             \\
$^{12}C$   & 0.68 & 0.11 &0.79 & 0.81 & 0.68                           \\
$^{16}O$   & 0.69  & 0.11 &0.80 & 0.83& 0.68                          \\
$^{40}Ca$  & 0.68 &0.10  &0.78 & 0.83 & 0.66                            \\
$^{56}Fe$  & 0.69 &0.10  &0.79 & 0.83 & 0.61                         \\
$^{208}Pb$ & 0.68 &0.13  &0.82 & 0.86 & 0.64                          \\
 \hline\hline
\end{tabular}
\label{tab32}
\end{table}
\newpage

\begin{table}[h]
\caption{The EMC ratio  (Eq. (\ref{true})) given by Eqs.
(\ref{erreaFS}), (\ref{errelanda})
 and (\ref{erreaVNC0}) multiplied by $A\,F_{2N}/F_{2D}$ (see text),  calculated at  $x=0.5$ with the value of
  $\lambda$
 fixed to reproduce the experimental data of $^{12}C$  ($\lambda(0.5)\approx 0.82$, for the case
 $\Delta E^{N/A}\neq \Delta E^{N/D}$, and  $\lambda(0.5)\approx 0.66$
 for $\Delta E_{N/A} = \Delta E^{N/D}=800\, MeV$, respectively). The theoretical results
  are compared with the SLAC experimental data
 fitted by
  ${R}^{exp}=1.009\,A^{-0.0234}$ \cite{Gomez}. Wave functions and Spectral functions
as in Tables \ref{tab1} and \ref{tab2}.\\
}
\begin{tabular}{|c|c|c|c|c||c|c|}\hline\hline
$A$  & ${R}_A^{LC(\delta)}$ &${R}_A^{LC(\delta)}$& ${R}_A^{LC}$ & ${R}_A^{VNC}$ & ${R}_A^{exp}$\\
       &$\Delta E^{N/A}\neq \Delta E^{N/D} $&$\Delta E^{N/A}= \Delta E^{N/D}$&&&\\
\hline %\tablenotemark[1]
$D$       &1.0   &1.0       &1.0 &1.0 &\\
$^3He  $  &0.99  &0.99      &1.0 &0.99 &0.980\\
$^4He  $  &0.96  &0.962     &1.0 &0.96 &0.980\\
$^{12}C$  &0.95  &0.95      &1.0 &0.955 &0.950\\
$^{16}O$  &0.95  &0.953     &1.0 &0.953 &0.946\\
$^{40}Ca$ &0.949 &0.948     &1.0 &0.952 &0.930\\
$^{56}Fe$ &0.951 &0.950     &1.0 &0.953 &0.920\\
$^{208}Pb$&0.943 &0.941     &1.0 &0.950 &0.890\\
 \hline\hline
\end{tabular}
\end{table}
\newpage
\end{document}